\documentclass[12pt]{article}

\def\eptwo{\left\{ \phantom{|}^{\mu\nu}_{ab} \right\}}
\def\epthree{\left\{ \phantom{|}^{\mu\nu\alpha}_{abc} \right\}}
\def\epfour{\left\{ \phantom{|}^{\mu\nu\alpha\beta}_{abcd} \right\}}
\def\epfive{\left\{ \phantom{|}^{\mu\nu\alpha\beta\gamma}_{abcde}%
\right\}}

\author{Yu.~M.~Zinoviev
       \thanks{E-mail address: Yurii.Zinoviev@ihep.ru} \\
        {\it Institute for High Energy Physics} \\
        {\it Protvino, Moscow Region, 142280, Russia}}
\title{Massive spin-2 in the Fradkin-Vasiliev formalism \\
       I. Partially massless case}

\date{}

\begin{document}

\maketitle

\begin{abstract}
We apply Fradkin-Vasiliev formalism to construction of non-trivial
cubic interaction vertices for massive spin-2 particles. In this first
paper as a relatively simple but instructive example we consider
self-interaction and gravitational interaction of partially massless
spin-2.
\end{abstract}

\thispagestyle{empty}
\newpage
\setcounter{page}{1}

\section*{Introduction}

Last years there were lot of activities in the investigation of
consistent cubic interaction vertices for higher spin fields. Such
investigations are very important steps in the search for consistent
higher spin theories and, in particular, provide information on the
possible gauge symmetry algebras behind such models. Till now most of
the results were devoted to cubic vertices for massless higher spin
fields and now we have rather good understanding of  their properties.
Moreover, the results obtained by different groups and different
methods are perfectly consistent, see e.g. \cite{Vas01}-\cite{HGR13}.

At the same time investigations of cubic vertices containing massive
higher spin fields are not so numerous, the most important one being
classification of cubic vertices in flat Minkowski space by Metsaev
\cite{Met05,Met07b,Met12}, while there also exist a number of
concrete examples e.g. \cite{Zin06}-\cite{BDT13}.

One of the approaches that turned out to be very effective for
investigation of massless higher spin fields interactions is the
Fradkin-Vasiliev formalism \cite{FV87,FV87a} (see also
\cite{Alk10,BSZ11,BS11,BPS12}). Let us briefly remind how this
formalism works. The basis for the whole construction is the
frame-like formalism \cite{Vas80,LV88,Vas88}, where massless higher
spin particle is described by a set of (physical, auxiliary and extra)
one forms that we will collectively denote as $\Phi$ here. As far as
the free theory is concerned, three most important facts are:
\begin{itemize}
\item each field has its own gauge transformation
$$
\delta \Phi \sim D \xi \oplus e  \xi
$$
where $D$ is $(A)dS$ covariant derivative, while $e$ is background
$(A)dS$ frame;
\item  gauge invariant two-form (curvature) can be
constructed for each field
$$
R \sim D \wedge \Phi \oplus e \wedge \Phi
$$
\item free Lagrangian can be rewritten in an explicitly gauge
invariant form
$$
{\cal L}_0 \sim \sum R \wedge R
$$
\end{itemize}
Using these ingredients cubic interaction vertices can be constructed
by the following straightforward steps.
\begin{itemize}
\item Take the most general quadratic deformation for curvatures
$$
\hat{R} = R \oplus \Phi \wedge \Phi
$$
as a result these new deformed curvatures ceased to be invariant
$$
\delta \hat{R} \sim \Phi \wedge D \xi \oplus e \wedge \Phi \xi
$$
\item Introduce corrections to gauge transformations 
$$
\delta \Phi \sim \Phi \xi
$$
in such a way that
$$
\delta \hat{R} \sim D \wedge \Phi \xi \oplus e \wedge \Phi \xi
$$
\item Adjust coefficients so that deformed curvatures transform
covariantly
$$
\delta \hat{R} \sim R \xi
$$
\item At last, consider the Lagrangian in the form
$$
{\cal L} \sim \sum \hat{R} \wedge \hat{R} \oplus \sum R \wedge R
\wedge \Phi
$$
where the first part is just the sum of free Lagrangians with initial
curvatures replaced by the deformed ones, while the second part
contains all possible abelian vertices. By construction and due to
Bianchi identities all variations for such Lagrangian take the form
$$
\delta {\cal L} \sim R \wedge R \xi
$$
reducing the problem to the set of algebraic equations. Moreover,
Vasiliev has shown \cite{Vas11} that for the three massless fields
with arbitrary spins $s_1$, $s_2$ and $s_3$ all non-trivial cubic
vertices having up to $s_1+s_2+s_3-2$ derivatives can be constructed
in this way.
\end{itemize}

As we have seen two main ingredients of this approach are frame-like
formalism and gauge invariance. But frame-like gauge invariant
description exists for massive higher spin fields as well
\cite{Zin08b}-\cite{PV10}. Thus it seems natural to
extend Fradkin-Vasiliev formalism to the cases where both massive and
(partially) massless fields are present. Such approach has already
been successfully applied to the investigation of gravitational and
electromagnetic interactions for simplest massive mixed symmetry field
\cite{Zin10a,Zin11}. Now we are going to apply this approach to the
construction of cubic vertices for massive spin-2
particles\footnote{Let us stress that we consider massive spin-2 as a
simple representative of massive higher spin fields, and not as
massive graviton.}. In this first paper we restrict ourselves with
relatively simple but instructive case of partially massless spin-2
field \cite{DW01,DW01a,DW01c,Zin01} leaving general massive case for
the second part.

The paper is organized as follows. In Section 1 we illustrate general
approach on the simplest but physically important example of massless
spin-2 field in $(A)dS$ space. Namely, we consider both
self-interaction as well as gravitational interaction vertices for
such field. The main section 2 is devoted to the partially massless
spin-2 case. First of all in subsection 2.1 we provide all necessary
kinematic formulas. Then in subsections 2.2 and 2.3 we consider
self-interaction and gravitational interaction correspondingly. Due to
the presence of zero forms as well as a number of identities making
different terms equivalent on-shell, an analysis turns out to be more
complicated than in the purely massless case. Thus as an
independent check for the number of non-equivalent cubic vertices (as
well as very instructive comparison) in Appendix we reconsider the
same vertices in a straightforward constructive approach.

\noindent
{\bf Notations and conventions} We work in $(A)dS$ space with
dimension $d \ge 4$ with (non-dynamical) background frame $e_\mu{}^a$
and $(A)dS$ covariant derivative $D_\mu$ normalized so that
$$
[ D_\mu, D_\nu ] \xi^a = - \kappa (e_\mu{}^a \xi_\nu - e_\nu{}^a
\xi_\mu), \qquad \kappa = \frac{2\Lambda}{(d-1)(d-2))}
$$
Here Greek letters are used for the world indices, while Latin letters
denote local ones. As it common for the frame-like formalism, all
terms in the Lagrangians will be completely antisymmetric on world
indices and we will heavily use notations like
$$
\eptwo = e^\mu{}_a e^\nu{}_b - e^\mu{}_b e^\nu{}_a
$$

\section{Massless case}

\subsection{Kinematics}

In the frame-like formalism free Lagrangian for massless spin 2 field
in $(A)dS$ background has the form:
\begin{equation}
{\cal L}_0 = \frac{1}{2} \eptwo \omega_\mu{}^{ac} \omega_\nu{}^{bc} -
\frac{1}{2} \epthree \omega_\mu{}^{ab} D_\nu h_\alpha{}^c -
\frac{(d-2)\kappa}{2} \eptwo h_\mu{}^a h_\nu{}^b
\end{equation}
This Lagrangian is invariant under the following gauge
transformations:
\begin{equation}
\delta_0 \omega_\mu{}^{ab} = D_\mu \hat{\eta}^{ab} + \kappa
e_\mu{}^{[a} \hat{\xi}^{b]}, \qquad \delta_0 h_\mu{}^a = D_\mu
\hat{\xi}^a + \hat{\eta}_\mu{}^a
\end{equation}
It is easy to construct two gauge invariant objects (linearized
curvature and torsion):
\begin{eqnarray}
R_{\mu\nu}{}^{ab} &=& D_{[\mu} \omega_{\nu]}{}^{ab} + \kappa 
e_{[\mu}{}^{[a} h_{\nu]}{}^{b]} \nonumber \\
T_{\mu\nu}{}^a &=& D_{[\mu} h_{\nu]}{}^a - \omega_{[\mu,\nu]}{}^a
\end{eqnarray}
Differential identities for them look like:
\begin{equation}
D_{[\mu} R_{\nu\alpha]}{}^{ab} = - \kappa e_{[\mu}{}^{[a}
T_{\nu\alpha]}{}^{b]}, \qquad
D_{[\mu} T_{\nu\alpha]}{}^a = - R_{[\mu\nu,\alpha]}{}^a
\end{equation}
Note that on mass shell for auxiliary field $\omega_\mu{}^{ab}$ we
have
$$
T_{\mu\nu}{}^a \approx 0 \quad \Rightarrow \quad 
R_{[\mu\nu,\alpha]}{}^a \approx 0, \qquad D_{[\mu} 
R_{\nu\alpha]}{}^{ab} \approx 0
$$
The free Lagrangian can be rewritten in the explicitly gauge
invariant form:
\begin{equation}
{\cal L}_0 = a_0 \epfour R_{\mu\nu}{}^{ab} R_{\alpha\beta}{}^{cd},
\qquad a_0 = - \frac{1}{32(d-3)\kappa}
\end{equation}

\subsection{Self-interaction}

The most general quadratic deformations for curvatures have the form:
\begin{eqnarray}
\Delta R_{\mu\nu}{}^{ab} &=& b_0 \omega_{[\mu}{}^{ca} 
\omega_{\nu]}{}^{bc} + b_1 h_{[\mu}{}^a h_{\nu]}{}^b \nonumber \\
\Delta T_{\mu\nu}{}^a &=& b_2 \omega_{[\mu}{}^{ab} h_{\nu]}{}^b
\end{eqnarray}
If we require that deformed curvatures transform covariantly we have
to put:
$$
b_1 = \kappa b_0, \qquad b_2 = b_0
$$
In this case corrections to the initial gauge transformations look
like
\begin{eqnarray}
\delta_1 \omega_\mu{}^{ab} &=& b_0 [ \omega_\mu{}^{c[a}
\hat{\eta}^{b]c} + \kappa h_\mu{}^{[a} \hat{\xi}^{b]} ] \nonumber \\
\delta_1 h_\mu{}^a &=& b_0 [ - \hat{\eta}^{ab} h_\mu{}^b +
\omega_\mu{}^{ab} \hat{\xi}^b ]
\end{eqnarray}
while deformed curvatures transform as follows:
\begin{eqnarray}
\delta \hat{R}_{\mu\nu}{}^{ab} &=& b_0 R_{\mu\nu}{}^{c[a}
\hat{\eta}^{b]c} + \kappa b_0 T_{\mu\nu}{}^{[a} \hat{\xi}^{b]}
\nonumber \\
\delta \hat{T}_{\mu\nu}{}^a &=& - b_0 \hat{\eta}^{ab} T_{\mu\nu}{}^b +
b_0 R_{\mu\nu}{}^{ab} \hat{\xi}^b
\end{eqnarray}
Let us consider the following Lagrangian
\begin{equation}
{\cal L} = a_0 \epfour \hat{R}_{\mu\nu}{}^{ab} 
\hat{R}_{\alpha\beta}{}^{cd} + c_0 \epfive R_{\mu\nu}{}^{ab}
R_{\alpha\beta}{}^{cd} h_\gamma{}^e
\end{equation}
where the first term is just the free Lagrangian where initial
curvature is replaced by the deformed one, while the second term is an
abelian vertex. Using identities given above it is easy to check that
both terms are gauge invariant on-shell. This Lagrangian gives
the following cubic vertex (here and in what follows the second index
denotes the number of derivatives in the vertex\footnote{Calculating
the number of derivatives we take into account that auxiliary field
$\omega_\mu{}^{ab}$ is equivalent to the first derivative of physical
field.}):
$$
{\cal L}_1 = {\cal L}_{14} + {\cal L}_{12} + {\cal L}_{10}
$$
\begin{eqnarray*}
{\cal L}_{14} &=& - 8a_0b_0 \epfour D_\mu \omega_\nu{}^{ab}
\omega_\alpha{}^{ce} \omega_\beta{}^{de} + 4c_0 \epfive D_\mu
\omega_\nu{}^{ab} D_\alpha \omega_\beta{}^{cd} h_\gamma{}^e \\
{\cal L}_{12} &=& 8\kappa (a_0b_0 + 2c_0(d-4)) \epfour D_\mu
\omega_\nu{}^{ab} h_\alpha{}^c h_\beta{}^d \\
 && - 16a_0b_0(d-3)\kappa \epthree h_\mu{}^a \omega_\nu{}^{bd}
\omega_\alpha{}^{cd} \\
{\cal L}_{10} &=& 16(d-3)\kappa^2 ( a_0b_0 + c_0(d-4)) \epthree
h_\mu{}^a h_\nu{}^b h_\alpha{}^c
\end{eqnarray*}
Thus we have two independent vertices with terms up to four
derivatives\footnote{We are working in the linear approximation so for
any two solutions their arbitrary linear combination is also a
solution. Thus the number of independent solutions is just the number
of free parameters.}. But on-shell we have
$$
4c_0 \epfive D_\mu \omega_\nu{}^{ab} D_\alpha \omega_\beta{}^{cd}
h_\gamma{}^e \approx 12c_0 \epfour D_\mu \omega_\nu{}^{ab}
\omega_\alpha{}^{ce} \omega_\beta{}^{de} -
$$
$$
- 8c_0(d-4)\kappa \epthree [ 2 \omega_\mu{}^{ad} \omega_\nu{}^{bd}
h_\alpha{}^c + \omega_\mu{}^{ab} \omega_\nu{}^{cd} h_\alpha{}^d ] 
$$
Thus if we put
$$
c_0 = \frac{2a_0b_0}{3}
$$
all four derivative terms cancel on-shell leaving us with the vertex
containing no more than two derivatives:
\begin{equation}
{\cal L}_1 = \frac{b_0}{2} [ \epthree \omega_\mu{}^{ad}
\omega_\nu{}^{bd} h_\alpha{}^c - \frac{1}{2} \epfour D_\mu
\omega_\nu{}^{ab} h_\alpha{}^c h_\beta{}^d - \frac{(2d-5)\kappa}{3}
\epthree h_\mu{}^a h_\nu{}^b h_\alpha{}^c ]
\end{equation}

\subsection{Gravitational interaction}

For the second spin-2 we will use notations ($\Omega_\mu{}^{ab}$,
$f_\mu{}^a$), ($\eta^{ab}$, $\xi^a$) and (${\cal F}_{\mu\nu}{}^{ab}$,
${\cal T}_{\mu\nu}{}^a$) for fields, gauge parameters and gauge
invariant curvatures correspondingly.

Let us consider gravitational interactions for this second spin-2.
Similarly to the previous case for the deformations of gravitational
curvatures we obtain
\begin{eqnarray}
\Delta R_{\mu\nu}{}^{ab} &=& b_0 [ \Omega_{[\mu}{}^{ca} 
\Omega_{\nu]}{}^{bc} + \kappa f_{[\mu}{}^a f_{\nu]}{}^b ] \nonumber \\
\Delta T_{\mu\nu}{}^a &=& b_0 \Omega_{[\mu}{}^{ab} f_{\nu]}{}^b
\end{eqnarray}
while deformed curvatures will transform as follows:
\begin{eqnarray}
\delta \hat{R}_{\mu\nu}{}^{ab} &=& b_0 {\cal F}_{\mu\nu}{}^{c[a}
\eta^{b]c} + \kappa b_0 {\cal T}_{\mu\nu}{}^{[a} \xi^{b]} \nonumber \\
\delta \hat{T}_{\mu\nu}{}^a &=& - b_0 \eta^{ab} {\cal T}_{\mu\nu}{}^b
+ b_0 {\cal F}_{\mu\nu}{}^{ab} \xi^b
\end{eqnarray}
As for the deformations for the second spin-2 curvatures they
correspond to standard minimal substitution rules for gravitational
interactions:
\begin{eqnarray}
\Delta {\cal F}_{\mu\nu}{}^{ab} &=& b_1 [ \Omega_{[\mu}{}^{c[a}
\omega_{\nu]}{}^{b]c} + \kappa f_{[\mu}{}^{[a} h_{\nu]}{}^{b]} ]
\nonumber \\
\Delta {\cal T}_{\mu\nu}{}^a &=& b_1 [ \Omega_{[\mu}{}^{ab} 
h_{\nu]}{}^b + \omega_{[\mu}{}^{ab} f_{\nu]}{}^b ]
\end{eqnarray}
while transformation rules for them look like:
\begin{eqnarray}
\delta \hat{\cal F}_{\mu\nu}{}^{ab} &=& b_1 [ - \hat{\eta}^{c[a} 
{\cal F}_{\mu\nu}{}^{b]c} + \kappa {\cal T}_{\mu\nu}{}^{[a} 
\hat{\xi}{}^{b]} - \eta^{c[a} R_{\mu\nu}{}^{b]c} + \kappa 
T_{\mu\nu}{}^{[a} \xi^{b]} ] \nonumber \\
\delta \hat{\cal T}_{\mu\nu}{}^a &=& b_1 [ - \hat{\eta}^{ab} 
{\cal T}_{\mu\nu}{}^b + {\cal F}_{\mu\nu}{}^{ab} \hat{\xi}^b -
\eta^{ab} T_{\mu\nu}{}^b + R_{\mu\nu}{}^{ab} \xi^b ]
\end{eqnarray}
Note that at this stage two parameters $b_0$ and $b_1$ are independent
and it may seem that it contradicts with the universality of
gravitational interactions. The reason is that covariance of deformed
curvatures guarantees that equation of motion for the theory we are
trying to construct will be gauge invariant but it does not guarantee
that these equations will be Lagrangean. Thus if we put these deformed
curvatures into the Lagrangian and require that this Lagrangian be
invariant we have to expect that parameters $b_0$ and $b_1$ will be
related. As we will see right now it turns out to be the case.

Let us consider the following Lagrangian:
\begin{equation}
{\cal L} = a_0 \epfour [ \hat{\cal F}_{\mu\nu}{}^{ab} 
\hat{\cal F}_{\alpha\beta}{}^{cd} + \hat{R}_{\mu\nu}{}^{ab}
\hat{R}_{\alpha\beta}{}^{cd} ] + c_1 \epfive  
{\cal F}_{\mu\nu}{}^{ab} {\cal F}_{\alpha\beta}{}^{cd} h_\gamma{}^e
\end{equation}
where the first two terms are just the sum of free Lagrangians with
initial curvatures replaced by the deformed ones, while the last term
is an abelian vertex. Note that there is one more abelian vertex
$$
\Delta {\cal L} = c_2 \epfive {\cal F}_{\mu\nu}{}^{ab} 
R_{\alpha\beta}{}^{cd} f_\gamma{}^e
$$
but (as we have explicitly checked) this vertex completely equivalent
on-shell to the one with coefficient $c_1$ so we will not introduce it
here. Let us take transformations of curvatures that do not vanish 
on-shell:
\begin{eqnarray}
\delta \hat{\cal F}_{\mu\nu}{}^{ab} &=& - b_1 [ \hat{\eta}^{c[a}
{\cal F}_{\mu\nu}{}^{b]c} + \eta^{c[a} R_{\mu\nu}{}^{b]c} ] \nonumber
\\
\delta \hat{R}_{\mu\nu}{}^{ab} &=& - b_0 \eta^{c[a} 
{\cal F}_{\mu\nu}{}^{b]c}
\end{eqnarray}
Variations under the $\hat{\eta}^{ab}$ transformations trivially
vanish on-shell, so let us consider the ones for the $\eta^{ab}$
transformations:
$$
- 4a_0 \epfour [ b_1 {\cal F}_{\mu\nu}{}^{ab} R_{\alpha\beta}{}^{ce}
\eta^{de} + b_0 {\cal F}_{\mu\nu}{}^{ae} R_{\alpha\beta}{}^{bc}
\eta^{de} ]
$$
But on-shell we have two identities
$$
0 \approx \epfive {\cal F}_{\mu\nu,\alpha}{}^a R_{\alpha\beta}{}^{bc}
\eta^{de} = 2\epfour {\cal F}_{\mu\nu}{}^{ae} [ - 
R_{\alpha\beta}{}^{be} \eta^{cd} - R_{\alpha\beta}{}^{bc} \eta^{de} ] 
$$
$$
0 \approx \epfive {\cal F}_{\mu\nu}{}^{ab} R_{\alpha\beta,\gamma}{}^c
\eta^{de} = 2 \epfour [ {\cal F}_{\mu\nu}{}^{ae} 
R_{\alpha\beta}{}^{be} \eta^{cd} - {\cal F}_{\mu\nu}{}^{ab}
R_{\alpha\beta}{}^{ce} \eta^{de} ]
$$
and their combination gives us
\begin{equation}
\epfour [ {\cal F}_{\mu\nu}{}^{ab} R_{\alpha\beta}{}^{ce} + 
{\cal F}_{\mu\nu}{}^{ae} R_{\alpha\beta}{}^{bc} ] \eta^{de} \approx 0
\label{id1}
\end{equation}
Thus we have to put (as expected)
\begin{equation}
b_1 = b_0
\end{equation}
So, as in the previous case, we have two independent vertices with
free parameters $b_0$ and $c_1$ and terms containing up to four
derivatives. Let us extract all the terms for the cubic vertex:
$$
{\cal L}_1 = {\cal L}_{14} + {\cal L}_{12} + {\cal L}_{10}
$$
\begin{eqnarray*}
{\cal L}_{14} &=& - 8a_0b_0 \epfour [ 2 D_\mu \Omega_\nu{}^{ab}
\Omega_\alpha{}^{ce} \omega_\beta{}^{de} + D_\mu \omega_\nu{}^{ab}
\Omega_\alpha{}^{ce} \Omega_\beta{}^{de} ] \\
 && + 4c_1 \epfive  D_\mu \Omega_\nu{}^{ab} D_\alpha
\Omega_\beta{}^{cd} h_\gamma{}^e \\
{\cal L}_{12} &=& - 8a_0b_0\kappa(d-3) \epthree [
\Omega_\mu{}^{ad} \Omega_\nu{}^{bd} + 2 \Omega_\mu{}^{ad}
\omega_\nu{}^{bd} f_\alpha{}^c ] \\
 && + 8\kappa \epfour [ (2a_0b_0+2(d-4)c_1) D_\mu \Omega_\nu{}^{ab}
f_\alpha{}^c h_\beta{}^d + a_0b_0 D_\mu \omega_\nu{}^{ab} f_\alpha{}^c
f_\beta{}^d ] \\
{\cal L}_{10} &=& 16\kappa^2(d-3) (3a_0b_0+(d-4)c_1) \epthree
f_\mu{}^a f_\nu{}^b h_\alpha{}^c
\end{eqnarray*}
But on the auxiliary fields $\omega_\mu{}^{ab}$ and
$\Omega_\mu{}^{ab}$ mass shell we have
\begin{eqnarray*}
4c_1 \epfive D_\mu \Omega_\nu{}^{ab} D_\alpha \Omega_\beta{}^{cd}
h_\gamma{}^e &\approx& 4c_1 \epfour [ 2 D_\mu \Omega_\nu{}^{ab}
\Omega_\alpha{}^{ce} \omega_\beta{}^{de} + D_\mu \omega_\nu{}^{ab}
\Omega_\alpha{}^{ce} \Omega_\beta{}^{de} ] \\
 && - 4\kappa(d-4)c_1 \epfour [ 2 D_\mu \Omega_\nu{}^{ab} f_\alpha{}^c
h_\beta{}^d - D_\mu \omega_\nu{}^{ab} f_\alpha{}^c f_\beta{}^d ]
\end{eqnarray*}
thus for $c_1 = 2a_0b_0$ all four derivative terms vanish leaving us
with the vertex containing no more than two derivatives:
\begin{eqnarray}
{\cal L}_1 &=& \frac{b_0}{4} \epthree [ \Omega_\mu{}^{ad}
\Omega_\nu{}^{bd} + 2 \Omega_\mu{}^{ad} \omega_\nu{}^{bd} f_\alpha{}^c
] \nonumber \\
 && - \frac{b_0}{4} \epfour [ 2 D_\mu \Omega_\nu{}^{ab}
f_\alpha{}^c h_\beta{}^d + D_\mu \omega_\nu{}^{ab} f_\alpha{}^c
f_\beta{}^d ] \nonumber \\
 && - \frac{(2d-5)\kappa b_0}{2} \epthree f_\mu{}^a f_\nu{}^b
h_\alpha{}^c
\end{eqnarray}

\section{Partially massless case}

\subsection{Kinematics}

In the frame-like formalism gauge invariant description for the
partially massless spin-2 particle \cite{Zin08b,PV10} requires two
pairs of (auxiliary and physical) fields:
$(\Omega_\mu{}^{ab},f_\mu{}^a)$ and $(B^{ab},B_\mu)$. Free Lagrangian
for such particle has the form:
\begin{eqnarray}
{\cal L}_0 &=& \frac{1}{2} \eptwo \Omega_\mu{}^{ac} \Omega_\nu{}^{bc}
- \frac{1}{2} \epthree \Omega_\mu{}^{ab} D_\nu f_\alpha{}^c + 
\frac{1}{2} B_{ab}{}^2 - \eptwo B^{ab} D_\mu B_\nu \nonumber \\
 && + m [ \eptwo \omega_\mu{}^{ab} B_\nu + e^\mu{}_a B^{ab} f_\mu{}^b
]
\end{eqnarray}
where $m^2 = (d-2)\kappa$. This Lagrangian is invariant under the
following gauge transformations:
\begin{eqnarray}
\delta_0 \Omega_\mu{}^{ab} &=& D_\mu \eta^{ab}, \qquad
\delta_0 f_\mu{}^a = D_\mu \xi^a + \eta_\mu{}^a + \frac{2m}{(d-2)}
e_\mu{}^a \xi \nonumber \\
\delta_0 B^{ab} &=& - m \eta^{ab}, \qquad
\delta_0 B_\mu = D_\mu \xi + \frac{m}{2} \xi_\mu
\end{eqnarray}
Correspondingly, we have four gauge invariant objects (curvatures):
\begin{eqnarray}
{\cal F}_{\mu\nu}{}^{ab} &=& D_{[\mu} \Omega_{\nu]}{}^{ab} -
\frac{m}{(d-2)} e_{[\mu}{}^{[a} B_{\nu]}{}^{b]} \nonumber \\
{\cal T}_{\mu\nu}{}^a &=& D_{[\mu} f_{\nu]}{}^a -
\Omega_{[\mu,\nu]}{}^a + \frac{2m}{(d-2)} e_{[\mu}{}^a B_{\nu]}
\nonumber \\
{\cal B}_\mu{}^{ab} &=& D_\mu B^{ab} + m \Omega_\mu{}^{ab} \\
{\cal B}_{\mu\nu} &=& D_{[\mu} B_{\nu]} - B_{\mu\nu} - \frac{m}{2}
f_{[\mu,\nu]} \nonumber
\end{eqnarray}
They satisfy the following differential identities:
\begin{eqnarray}
D_{[\mu} {\cal F}_{\nu\alpha]}{}^{ab} &=& \frac{m}{(d-2)} 
e_{[\mu}{}^{[a} {\cal B}_{\nu,\alpha]}{}^{b]} \nonumber \\
D_{[\mu} {\cal T}_{\nu\alpha]}{}^a &=& - 
{\cal F}_{[\mu\nu,\alpha]}{}^a - \frac{2m}{(d-2)} e_{[\mu}{}^a 
{\cal B}_{\nu\alpha]} \nonumber \\
D_{[\mu} {\cal B}_{\nu]}{}^{ab} &=& m {\cal F}_{\mu\nu}{}^{ab} \\
D_{[\mu} {\cal B}_{\nu\alpha]} &=& - {\cal B}_{[\mu,\nu\alpha]} -
\frac{m}{2} {\cal T}_{[\mu\nu,\alpha]} \nonumber
\end{eqnarray}
Note that on mass shell for auxiliary fields $\Omega_\mu{}^{ab}$ and
$B^{ab}$ we have
$$
{\cal T}_{\mu\nu}{}^a \approx 0, \quad {\cal B}_{\mu\nu} \approx 0
\quad \Rightarrow \quad {\cal F}_{[\mu\nu,\alpha]}{}^a \approx 0,
\quad {\cal B}_{[\mu,\nu\alpha]} \approx 0
$$

Using these curvatures the free Lagrangian can be rewritten in an
explicitly gauge invariant form
\begin{equation}
{\cal L}_0 = a_1 \epfour {\cal F}_{\mu\nu}{}^{ab} 
{\cal F}_{\alpha\beta}{}^{cd} + a_2 \eptwo {\cal B}_\mu{}^{ac} 
{\cal B}_\nu{}^{bc} + a_3 \epthree {\cal B}_\mu{}^{ab} 
{\cal T}_{\nu\alpha}{}^c
\end{equation}
where
$$
\frac{16(d-3)}{(d-2)} a_1 - a_2 = \frac{1}{2m^2}, \qquad 
a_3 = - \frac{1}{4m}
$$
The ambiguity with coefficients is related with the identity
$$
0 = \epfour D_\mu [ {\cal F}_{\nu\alpha}{}^{ab} {\cal B}_\beta{}^{cd}
] = \epfour [ {\cal F}_{\mu\nu}{}^{ab} D_\alpha {\cal B}_\beta{}^{cd}
+ D_\mu {\cal F}_{\nu\alpha}{}^{ab} {\cal B}_\beta{}^{cd} ]
$$
$$
= \frac{m}{2} \epfour {\cal F}_{\mu\nu}{}^{ab} 
{\cal F}_{\alpha\beta}{}^{cd} + \frac{8m(d-3)}{(d-2)} \eptwo
{\cal B}_\mu{}^{ac} {\cal B}_\nu{}^{bc} 
$$
In what follows we will use the convenient choice
\begin{equation}
a_1 = - \frac{(d-2)}{32(d-3)m^2}, \qquad a_2 = - \frac{1}{m^2}, \qquad
a_3 = - \frac{1}{4m}
\end{equation}

\subsection{Self-interaction}

Following general procedure we begin with the most general quadratic
deformations for all four curvatures:
\begin{eqnarray*}
\Delta {\cal F}_{\mu\nu}{}^{ab} &=& d_1 \Omega_{[\mu}{}^{c[a}
\Omega_{\nu]}{}^{b]c} + d_2 B_{[\mu}{}^{[a} B_{\nu]}{}^{b]} + d_3
B^{ab} B_{\mu\nu} + d_4 e_{[\mu}{}^{[a} B^{b]c} B_{\nu]}{}^c + d_5 
e_{[\mu}{}^a e_{\nu]}{}^b B^{cd} B^{cd}  \\
 && + d_6 \Omega_{[\mu}{}^{ab} B_{\nu]} + d_7 B_{[\mu}{}^{[a} 
f_{\nu]}{}^{b]} + d_8 B^{ab} f_{[\mu,\nu]} + d_9 e_{[\mu}{}^{[a}
B^{b]c} f_{\nu]}{}^c + d_{10} f_{[\mu}{}^{[a} f_{\nu]}{}^{b]} \\
\Delta {\cal T}_{\mu\nu}{}^a &=& d_{11} \Omega_{[\mu}{}^{ab} 
B_{\nu]}{}^b + d_{12} B^{ab} \Omega_{[\mu,\nu]}{}^b + d_{13} 
e_{[\mu}{}^a \Omega_{\nu]}{}^{bc} B^{bc} \\
 && + d_{14} \Omega_{[\mu}{}^{ab} f_{\nu]}{}^b + d_{15} B_{[\mu}{}^a
B_{\nu]} + d_{16} f_{[\mu}{}^a B_{\nu]} \\
\Delta {\cal B}_\mu{}^{ab} &=& d_{17} \Omega_\mu{}^{c[a} B^{b]c} +
d_{18} B^{ab} B_\mu \\
\Delta {\cal B}_{\mu\nu} &=& d_{19} B_{[\mu}{}^a f_{\nu]}{}^a
\end{eqnarray*}
Usual requirement that deformed curvatures transform covariantly gives
solution with five arbitrary parameters. However, due to the presence
of zero form $B^{ab}$ there are four possible field redefinitions
(their explicit action can be seen in the Appendix):
$$
\Omega_\mu{}^{ab} \Rightarrow \Omega_\mu{}^{ab} + \kappa_1 B^{ab}
B_\mu, \qquad
f_\mu{}^a \Rightarrow f_\mu{}^a + \kappa_2 B^{ab} f_\mu{}^b + \kappa_3
B^{ab} B_\mu{}^b + \kappa_4 e_\mu{}^a B^{bc} B^{bc}
$$
Using this freedom we can bring the deformations into the form
\begin{eqnarray}
\Delta {\cal F}_{\mu\nu}{}^{ab} &=& d_1 [ \Omega_{[\mu}{}^{c[a}
\Omega_{\nu]}{}^{b]c} + \frac{1}{(d-2)} ( B_{[\mu}{}^{[a} 
B_{\nu]}{}^{b]} - e_{[\mu}{}^{[a} B^{b]c} B_{\nu]}{}^c ) ] + \nonumber
\\
 && + d_6 [ \Omega_{[\mu}{}^{ab} B_{\nu]} - \frac{1}{m} B^{ab}
B_{\mu\nu} - \frac{1}{2} B^{ab} f_{[\mu,\nu]} ] \nonumber \\
\Delta {\cal T}_{\mu\nu}{}^a &=& 2d_1 \Omega_{[\mu}{}^{ab}
f_{\nu]}{}^b \\
\Delta {\cal B}_\mu{}^{ab} &=& d_1 \Omega_{[\mu}{}^{c[a} B^{b]c} -
d_6 B^{ab} B_\mu \nonumber \\
\Delta {\cal B}_{\mu\nu} &=& - d_1 B_{[\mu}{}^a f_{\nu]}{}^a \nonumber
\end{eqnarray}
where
$$
d_6 = - \frac{4md_1}{(d-2)}
$$
This corresponds to the following gauge transformations for the
deformed curvatures:
\begin{eqnarray}
\delta \hat{\cal F}_{\mu\nu}{}^{ab} &=& 2d_1 {\cal F}_{\mu\nu}{}^{c[a}
\eta^{b]c} + \frac{d_6}{2} {\cal B}_{[\mu}{}^{ab} \xi_{\nu]} + d_6
{\cal F}_{\mu\nu}{}^{ab} \xi \nonumber \\
\delta \hat{\cal T}_{\mu\nu}{}^a &=& - 2d_1 \eta^{ab} 
{\cal T}_{\mu\nu}{}^b + 2d_1 {\cal F}_{\mu\nu}{}^{ab} \xi^b \nonumber
\\
\delta \hat{\cal B}_\mu{}^{ab} &=& - d_1 \eta^{c[a} 
{\cal B}_\mu{}^{b]c} + d_6 {\cal B}_\mu{}^{ab} \xi \\
\delta \hat{\cal B}_{\mu\nu} &=&  - d_1 {\cal B}_{[\mu,\nu]}{}^a \xi^a
\nonumber
\end{eqnarray}

Now let us consider the following Lagrangian:
\begin{eqnarray}
{\cal L} &=& a_1 \epfour \hat{\cal F}_{\mu\nu}{}^{ab} 
\hat{\cal F}_{\alpha\beta}{}^{cd} + a_2 \eptwo \hat{\cal B}_\mu{}^{ac}
\hat{\cal B}_\nu{}^{bc} + a_3 \epthree \hat{\cal B}_\mu{}^{ab} 
\hat{\cal T}_{\nu\alpha}{}^c \nonumber \\
 && + a_4 \epfive {\cal F}_{\mu\nu}{}^{ab} 
{\cal F}_{\alpha\beta}{}^{cd} f_\gamma{}^e + a_5 \epthree 
{\cal B}_\mu{}^{ad} {\cal B}_\nu{}^{bd} f_\alpha{}^c + a_6 \epfour
{\cal F}_{\mu\nu}{}^{ab} {\cal B}_\alpha{}^{cd} B_\beta
\end{eqnarray}
Here the first line is just the free Lagrangian where initial
curvatures are replaced by the deformed ones, while the second line
contains possible abelian vertices\footnote{Note that in the partially
massless case (and in the massive case too) due to peculiarities of
gauge transformations the terms in the second line are not gauge
invariant separately.}.

Let us require that this Lagrangian be gauge invariant. All
$\eta^{ab}$ variations vanish on-shell. For the $\xi$ variations
we obtain
$$
(2d_6a_1 + \frac{2(d-4)ma_4}{(d-2)} + \frac{ma_6}{2}) \epfour 
{\cal F}_{\mu\nu}{}^{ab} {\cal F}_{\alpha\beta}{}^{cd} \xi
$$
$$
+ (2d_6a_2+2ma_5 + \frac{8(d-3)ma_6}{(d-2)}) \eptwo 
{\cal B}_\mu{}^{ac} {\cal B}_\nu{}^{bc} \xi = 0
$$
Thus we have to put
\begin{equation}
2d_6a_1 + \frac{2(d-4)ma_4}{(d-2)} + \frac{ma_6}{2} = 0
\end{equation}
\begin{equation}
2d_6a_2+2ma_5 + \frac{8(d-3)ma_6}{(d-2)} = 0
\end{equation}
For the $\xi^a$ transformations we get
$$
(4d_6a_1+ma_6+2d_1a_3) \epthree  {\cal F}_{\mu\nu}{}^{ad} 
{\cal B}_\alpha{}^{bc} \xi^d 
$$
$$
+ (\frac{16(d-4)ma_4}{(d-2)} - ma_5) \epthree  
{\cal F}_{\mu\nu}{}^{ad} {\cal B}_\alpha{}^{bd} \xi^c
$$
and using on-shell identity
\begin{equation}
0 \approx\epfour {\cal F}_{\mu\nu,\alpha}{}^a {\cal B}_\beta{}^{bc}
\xi^d = \epthree {\cal F}_{\mu\nu}{}^{ad} [ - 2 {\cal B}_\alpha{}^{bd}
\xi^c + {\cal B}_\alpha{}^{bc} \xi^d ] \label{id2}
\end{equation}
we obtain
\begin{equation}
8d_6a_1 + 2ma_6 + 4d_1a_3 + \frac{16(d-4)ma_4}{(d-2)} - ma_5 = 0
\end{equation}

Thus we obtain three equations which uniquely determines all free
coefficients $a_{4,5,6}$ so we have one cubic vertex with terms up to
four derivatives. Note that the case $d=4$ is special because the term
with coefficient $a_4$ is absent. Happily the solution still exists,
namely
$$
a_5 = - \frac{d_1}{m^2}, \qquad a_6 = - \frac{d_1}{2m^2}
$$
Moreover, as we have explicitly checked, all cubic terms with four
and three derivatives vanish on-shell and we reproduce rather well
known two derivative vertex \cite{Zin06,DJW13,RHRT13}. Note also that
the same general results (one four derivative vertex in $d > 4$ and
one two derivative vertex in $d=4$) was obtained also in \cite{JLT12}.

\subsection{Gravitational interaction}

We begin with the most general quadratic deformations for
gravitational curvatures:
\begin{eqnarray}
\Delta R_{\mu\nu}{}^{ab} &=& b_1 \Omega_{[\mu}{}^{c[a} 
\Omega_{\nu]}{}^{b]c} + b_2 B_{[\mu}{}^{[a} B_{\nu]}{}^{b]} + b_3
B_{\mu\nu} B^{ab} + b_4 e_{[\mu}{}^{[a} B^{b]c} B_{\nu]}{}^c + b_5
e_{[\mu}{}^a e_{\nu]}{}^b B^{cd} B^{cd} \nonumber \\
 && + b_6 \Omega_{[\mu}{}^{ab} B_{\nu]} + b_7 B_{[\mu}{}^{[a} 
f_{\nu]}{}^{b]} + b_8 B^{ab} f_{[\mu,\nu]} + b_9 e_{[\mu}{}^{[a}
B^{b]c} f_{\nu]}{}^c + b_{10} f_{[\mu}{}^{[a} f_{\nu]}{}^{b]} \\
\Delta T_{\mu\nu}{}^a &=& b_{11} \Omega_{[\mu}{}^{ab} f_{\nu]}{}^b +
b_{12} f_{[\mu}{}^a B_{\nu]} \nonumber
\end{eqnarray}
The solution has the form (again using all possible field
redefinitions):
$$
b_2 = - \frac{b_6}{4m}, \qquad
b_3 = - \frac{b_6}{m}, \qquad 
b_4 = - \frac{2b_1}{(d-2)}, \qquad
b_5 = 0
$$
$$
b_7 = - \frac{2mb_1}{(d-2)} - \frac{b_6}{2}, \qquad
b_8 = - \frac{b_6}{2}, \qquad 
b_9 = - \frac{2mb_1}{(d-2)}
$$
$$
b_{10} = - \frac{m^2b_1}{(d-2)} - \frac{mb_6}{4}, \qquad 
b_{11} = 2b_1, \qquad
b_{12} = \frac{4mb_1}{(d-2)} + b_6
$$
Gauge transformations for the deformed curvatures look like:
\begin{eqnarray}
\delta \hat{R}_{\mu\nu}{}^{ab} &=& 2b_1 {\cal F}_{\mu\nu}{}^{c[a}
\eta^{b]c} - b_6 {\cal B}_{\mu\nu} \eta^{ab} + b_7 
{\cal B}_{[\mu,\nu]}{}^{[a} \xi^{b]} - b_8 {\cal B}_{[\mu}{}^{ab}
\xi_{\nu]} - b_9 e_{[\mu}{}^{[a} {\cal B}_{\nu]}{}^{b]c} \xi^c
\nonumber \\
 &&  + 2b_{10} {\cal T}_{\mu\nu}{}^{[a} \xi^{b]} + b_6 
{\cal F}_{\mu\nu}{}^{ab} \xi \\
\delta \hat{T}_{\mu\nu}{}^a &=& - 2b_1 \eta^{ab} {\cal T}_{\mu\nu}{}^b
+ 2b_1 {\cal F}_{\mu\nu}{}^{ab} \xi^b - b_{12} {\cal B}_{\mu\nu} \xi^a
+ b_{12} {\cal T}_{\mu\nu}{}^a \xi \nonumber
\end{eqnarray}
As for the partially massless curvatures deformations they again
correspond to the minimal substitution rules:
\begin{eqnarray}
\Delta {\cal F}_{\mu\nu}{}^{ab} &=& - b_0 \omega_{[\mu}{}^{c[a}
\Omega_{\nu]}{}^{b]c} + \frac{mb_0}{(d-2)} [ B_{[\mu}{}^{[a} 
h_{\nu]}{}^{b]} - e_{[\mu}{}^{[a} B^{b]c} h_{\nu]}{}^c ] \nonumber \\
\Delta {\cal T}_{\mu\nu}{}^a &=& - b_0 \omega_{[\mu}{}^{ab} 
f_{\nu]}{}^b - b_0 \Omega_{[\mu}{}^{ab} h_{\nu]}{}^b - 
\frac{2mb_0}{(d-2)}  h_{[\mu}{}^a B_{\nu]} \nonumber \\
\Delta {\cal B}_\mu{}^{ab} &=& - b_0 \omega_\mu{}^{c[a} B^{b]c} \\
\Delta {\cal B}_{\mu\nu} &=& b_0 B_{[\mu}{}^a h_{\nu]}{}^a + 
\frac{mb_0}{2} f_{[\mu}{}^a h_{\nu]}{}^a \nonumber
\end{eqnarray}
while their transformations have the form:
\begin{eqnarray}
\delta \hat{\cal F}_{\mu\nu}{}^{ab} &=& - b_0 
{\cal F}_{\mu\nu}{}^{c[a} \hat{\eta}^{b]c} - b_0 R_{\mu\nu}{}^{c[a}
\eta^{b]c} + \frac{mb_0}{(d-2)} [ {\cal B}_{[\mu,\nu]}{}^{[a}
\hat{\xi}^{b]} + e_{[\mu}{}^{[a} {\cal B}_{\nu]}{}^{b]c} \hat{\xi}^c ]
\nonumber \\
\delta \hat{\cal T}_{\mu\nu}{}^a &=& b_0 \hat{\eta}^{ab} 
{\cal T}_{\mu\nu}{}^b - b_0 {\cal F}_{\mu\nu}{}^{ab} \hat{\xi}^b +
\frac{2mb_0}{(d-2)} {\cal B}_{\mu\nu} \hat{\xi}^a \nonumber \\
 && + b_0 \eta^{ab} T_{\mu\nu}{}^b - b_0 R_{\mu\nu}{}^{ab} \xi^b - 
\frac{2mb_0}{(d-2)} T_{\mu\nu}{}^a \xi \\
\delta \hat{\cal B}_\mu{}^{ab} &=& - b_0 {\cal B}_\mu{}^{c[a}
\hat{\eta}^{b]c} \nonumber \\
\delta \hat{\cal B}_{\mu\nu} &=& b_0 {\cal B}_{[\mu,\nu]}{}^a
\hat{\xi}^a + \frac{mb_0}{2} {\cal F}_{\mu\nu}{}^a \hat{\xi}^a -
\frac{mb_0}{2} T_{\mu\nu}{}^a \xi^a \nonumber
\end{eqnarray}

Now let us consider the following Lagrangian:
\begin{eqnarray}
{\cal L} &=& a_1 \epfour \hat{\cal F}_{\mu\nu}{}^{ab} 
\hat{\cal F}_{\alpha\beta}{}^{cd} + a_2 \eptwo \hat{\cal B}_\mu{}^{ac}
\hat{\cal B}_\nu{}^{bc} + a_3 \epthree \hat{\cal B}_\mu{}^{ab} 
\hat{\cal T}_{\nu\alpha}{}^c + a_0 \epfour \hat{R}_{\mu\nu}{}^{ab}
\hat{R}_{\alpha\beta}{}^{cd} \nonumber \\
 && + a_4 \epfive {\cal F}_{\mu\nu}{}^{ab} 
{\cal F}_{\alpha\beta}{}^{cd} h_\gamma{}^e + a_5 \epthree 
{\cal B}_\mu{}^{ad} {\cal B}_\nu{}^{bd} h_\alpha{}^c + a_6 \epfour
R_{\mu\nu}{}^{ab} {\cal B}_\alpha{}^{cd} B_\beta
\end{eqnarray}
Here the first line is the sum of the free Lagrangian for partially
massless and massless spin-2 where initial curvatures are replaced by
the deformed ones, while the second line contains possible abelian
vertices. Note that one more possible term
$$
\epfive {\cal F}_{\mu\nu}{}^{ab} R_{\alpha\beta}{}^{cd}
f_\gamma{}^e
$$
is on shell equivalent to some combination of others so we will not
introduce it here.

Now let us require that this Lagrangian be gauge invariant. All
$\hat{\eta}^{ab}$ variations vanish on-shell so we begin with
$\hat{\xi}^a$ transformations. We have to take into account the part
of variations that do not vanish on-shell, namely
$$
\delta \hat{\cal F}_{\mu\nu}{}^{ab} = \frac{mb_0}{(d-2)} 
{\cal B}_{[\mu,\nu]}{}^{[a} \hat{\xi}^{b]}, \qquad
\delta \hat{\cal T}_{\mu\nu}{}^a = - b_0 {\cal F}_{\mu\nu}{}^{ab}
\hat{\xi}^b 
$$
This produces:
$$
[ \frac{16m[(d-4)a_4-a_1b_0]}{(d-2)} - ma_5 ] \epthree 
{\cal F}_{\mu\nu}{}^{ad} {\cal B}_\alpha{}^{bd} \hat{\xi}^c  - a_3b_0
\epthree {\cal F}_{\mu\nu}{}^{ad} {\cal B}_\alpha{}^{bc} \hat{\xi}^d 
$$
Using on-shell identity
$$
0 \approx \epfour {\cal F}_{\mu\nu,\alpha}{}^a {\cal B}_\beta{}^{bc}
\hat{\xi}^d = \epthree {\cal F}_{\mu\nu}{}^{ad} [ - 2 
{\cal B}_\alpha{}^{bd} \hat{\xi}^c + {\cal B}_\alpha{}^{bc}
\hat{\xi}^d ] 
$$
we obtain first equation:
\begin{equation}
\frac{16[(d-4)a_4-a_1b_0]}{(d-2)} - a_5 + \frac{b_0}{2m^2} = 0
\end{equation}
For the $\eta^{ab}$ transformations we have to take into account only
$$
\delta \hat{R}_{\mu\nu}{}^{ab} = 2b_1 {\cal F}_{\mu\nu}{}^{c[a}
\eta^{b]c}, \qquad \delta \hat{\cal F}_{\mu\nu}{}^{ab} = - b_0
R_{\mu\nu}{}^{c[a} \eta^{b]c}
$$
This gives us
$$
4a_1b_0 \epfour {\cal F}_{\mu\nu}{}^{ab} R_{\alpha\beta}{}^{ce}
\eta^{de} - 8a_0b_1 \epfour {\cal F}_{\mu\nu}{}^{ae} \eta^{be} 
R_{\alpha\beta}{}^{cd}
$$
using once again on-shell identity (\ref{id1}) we obtain
\begin{equation}
a_1b_0 + 2a_0b_1 = 0
\end{equation}
Non-vanishing on-shell part of the $\xi^a$ transformations has the
form
$$
\delta \hat{R}_{\mu\nu}{}^{ab} = b_7 {\cal B}_{[\mu,\nu]}{}^{[a}
\xi^{b]} - b_8 {\cal B}_{[\mu}{}^{ab} \xi_{\nu]}, \qquad
\delta \hat{\cal T}_{\mu\nu}{}^a = - b_0 R_{\mu\nu}{}^{ab} \xi^b
$$
and we get
$$
- (a_3b_0+8a_0b_8-ma_6) \epthree R_{\mu\nu}{}^{ad} 
{\cal B}_\alpha{}^{bc} \xi^d - 16a_0b_7 \epthree R_{\mu\nu}{}^{ad}
{\cal B}_\alpha{}^{bd} \xi^c 
$$
Using on-shell identity (\ref{id2}) (where ${\cal F}_{\mu\nu}{}^{ab}$
is replaced by $R_{\mu\nu}{}^{ab}$) we obtain
\begin{equation}
a_3b_0+8a_0b_8-ma_6 + 8a_0b_7 = 0
\end{equation}
At last let us consider variations under $\xi$ transformations:
$$
\delta \hat{R}_{\mu\nu}{}^{ab} = b_6 {\cal F}_{\mu\nu}{}^{ab} \xi
$$
This produces
$$
[ 2a_0b_6 + \frac{ma_6}{2} ] \epfour {\cal F}_{\mu\nu}{}^{ab}
R_{\alpha\beta}{}^{cd} \xi
$$
and we obtain the last equation
\begin{equation}
4a_0b_6 + ma_6 = 0
\end{equation}
These equations have the following solution
\begin{equation}
b_1 = - \frac{b_0}{2}, \qquad b_6 = 2mb_0, \qquad
a_5 = \frac{16(d-4)a_4}{(d-2)} - 16a_0b_0, \qquad a_6 = - 8a_0b_0
\end{equation}
Thus in general $d > 4$ case we have two independent vertices with
parameters $b_0$ and $a_4$\footnote{As it will be shown in the
Appendix in $d=3$ case there exists one more cubic vertex with no more
than two derivatives, but Fradkin-Vasiliev formalism we use here works
in  $d \ge 4$ dimensions only so we did not obtain such vertex here.
Note also that in a frame-like gauge invariant formalism this vertex
has been constructed in \cite{Zin12}.}. In $d=4$ the parameter
$a_4$ is absent leaving with one vertex only. Moreover we have
explicitly checked that in this case all four derivative terms vanish
on-shell. Note that, contrary to the self-interaction case, here our
results do not agree with the one obtained in \cite{JLT12}. Table
"2-2-2 couplings" in Appendix B of this paper gives four
non-trivial vertices: two four derivatives ones and two vertices
having no more than two derivatives. Moreover these results do not
depend on space-time dimension. Due to very different approach used
by authors of \cite{JLT12} it is not an easy task to see where and why
such difference arises.

\section*{Conclusion}

As we have seen application of Fradkin-Vasiliev formalism to the
partially massless (and even more so in the massive) case appears to
be more complicated and less elegant. The reason is that due to the
large number of fields (main and Stueckelberg) and due to the presence
of zero forms one faces a lot of ambiguities related with non-trivial
on-shell identities and field redefinitions. Nevertheless, the
formalism does work and allows one to obtain reasonable results.

\vskip 1cm \noindent
{\bf Acknowledgment} Author is grateful to R.~R.~Metsaev and
E.~D.~Skvortsov for useful discussions. The work was supported in
parts by RFBR grant No.14-02-01172.

\appendix

\section{Partially massless spin-2 in a constructive approach}

In this appendix as an independent check for the results obtained in
the main part we reconsider the same problems in the straightforward
constructive approach.

\subsection{Modified 1 and $\frac{1}{2}$ order formalism}

In the constructive approach one usually assumes that the action can
be considered as a row in the number of fields:
$$
S = S_0 + S_1 + S_2 + \dots
$$
where $S_0$ is free (quadratic) action, $S_1$ contains cubic terms,
$S_2$ --- quartic ones and so on. Similarly for the gauge
transformations one assumes:
$$
\delta \Phi = \delta_0 \Phi + \delta_1 \Phi + \delta_2 \Phi + \dots
$$
where $\delta_0$ is non-homogeneous part, $\delta_1$ is linear in
fields and so on. Then variations of the action under any gauge
transformations can also be represented as a row:
$$
\delta S = \frac{\delta S_0}{\delta \Phi} \delta_0 \Phi + 
(\frac{\delta S_1}{\delta \Phi} \delta_0 \Phi +
\frac{\delta S_0}{\delta \Phi} \delta_1 \Phi) + \dots
$$
First term simply implies that the free action $S_0$ must be gauge
invariant under the initial gauge transformations $\delta_0 \Phi$.
Thus the first non-trivial level (that we will call linear
approximation) looks as:
$$
\frac{\delta S_1}{\delta \Phi} \delta_0\Phi +
\frac{\delta S_0}{\delta \Phi} \delta_1 \Phi = 0
$$
Working with the frame-like formalism it is convenient to separate
physical $\Phi$ and auxiliary $\Omega$ fields. Than in the honest
first order formalism one has to achieve:
$$
\frac{\delta S_1}{\delta \Phi} \delta_0\Phi +
\frac{\delta S_1}{\delta \Omega} \delta_0 \Omega +
\frac{\delta S_0}{\delta \Phi} \delta_1 \Phi +
\frac{\delta S_0}{\delta \Omega} \delta_1 \Omega = 0
$$
It means that one has to consider the most general ansatz both the
cubic vertex as well for the corrections to gauge transformations for
the fields $\Phi$ and $\Omega$. Taking into account that equations for
auxiliary fields are algebraic and on their mass shell these fields
are equivalent to the derivatives of physical ones, in supergravities
there appeared a so-called 1 and $\frac{1}{2}$ order formalism.
Schematically it looks like:
$$
\left[ \frac{\delta S_1}{\delta \Phi} \delta_0\Phi +
\frac{\delta S_0}{\delta \Phi} \delta_1 \Phi 
\right]_{\frac{\delta (S_0+S_1)}{\delta \Omega} = 0} = 0
$$
So one needs the most general ansatz for cubic vertex and physical
fields gauge transformations only, but all calculations have to be
done up to the terms proportional to the auxiliary fields equations,
i.e. on their mass shell. Such approach turned out to be very
effective, but it requires explicit solution of non-linear equations
for auxiliary fields that can be rather complicated task. If we
restrict ourselves with the linear approximation than there exists one
more possibility that we will call modified 1 and $\frac{1}{2}$ order
formalism. It looks like:
$$
\left[ \frac{\delta S_1}{\delta \Phi} \delta_0\Phi +
\frac{\delta S_1}{\delta \Omega} \delta_0 \Omega +
\frac{\delta S_0}{\delta \Phi} \delta_1 \Phi 
\right]_{\frac{\delta S_0}{\delta \Omega} = 0} = 0
$$
The main achievements here are twofold. At first, we have not consider
the most general ansatz for cubic vertex but terms that are
non-equivalent on auxiliary fields mass shell only. At second, we need
explicit solution for the free auxiliary fields equations only. In
what follows we will use such modified formalism.

\subsection{Self-interaction}

Our aim here is to determine the number of independent vertices so to
simplify calculations in this and subsequent subsections we will
heavily use all possible field redefinitions and all existing
on-shell
identities. We will work in a up-down approach i.e. we begin with
four derivative terms, then we consider terms with three derivatives
and so on.

\noindent
{\bf Vertex $2-2-2$ with four derivatives} The only (on-shell
non-trivial) possibility here is:
$$
{\cal L}_{14a} = a_0 \epfour D_\mu \Omega_\nu{}^{ab}
\Omega_\alpha{}^{ce} \Omega_\beta{}^{de}
$$

\noindent
{\bf Vertex $2-1-1$ with four derivatives} The most
general\footnote{Up to the terms that are equivalent on-shell} ansatz
looks like:
\begin{eqnarray*}
{\cal L}_{14b} &=& \eptwo [ a_1 \Omega_\mu{}^{ab} D_\nu B^{cd} B^{cd}
+ a_2 \Omega_\mu{}^{ac} D_\nu B^{bd} B^{cd} + a_3 \Omega_\mu{}^{ac}
D_\nu B^{cd} B^{bd} \\
 && \qquad + a_4 \Omega_\mu{}^{cd} D_\nu B^{ab} B^{cd} + a_5
\Omega_\mu{}^{cd} D_\nu B^{ac} B^{bd} + a_6 \Omega_\mu{}^{cd} D_\nu
B^{cd} B^{ab} ]
\end{eqnarray*}
But we have three possible field redefinitions:
\begin{eqnarray*}
f_\mu{}^a &\Rightarrow& f_\mu{}^a + \kappa_1 B^{ab} B_\mu{}^a +
\kappa_2 e_\mu{}^a B^2 \\
B_\mu &\Rightarrow& B_\mu + \kappa_3 \Omega_\mu{}^{ab} B^{ab}
\end{eqnarray*}
and two on-shell identities (up to lower derivative terms):
\begin{eqnarray*}
0 &\approx& \epthree D_\mu \Omega_{\nu,\alpha}{}^d B^{ab} B^{cd} =
\eptwo D_\mu \Omega_\nu{}^{cd} [ B^{ab} B^{cd} - 2 B^{ac} B^{bd} ] \\
 &=& \eptwo \Omega_\mu{}^{cd} [ D_\nu B^{ab} B^{cd} - 4 D_\mu B^{ac}
B^{bd} + D_\nu B^{cd} B^{ab} ] \\
0 &\approx& \epfour \Omega_\mu{}^{ab} D_\nu B_{\alpha\beta} B^{cd} = 2
\eptwo [ \Omega_\mu{}^{ab} B^{cd} - 4 \Omega_\mu{}^{ac} B^{bd} +
\Omega_\mu{}^{cd} B^{ab} ] D_\nu B^{cd}
\end{eqnarray*}
Thus we have one independent vertex only in agreement with fact that
there exists only one cubic $2-1-1$ vertex with three derivatives for
the massless fields. In what follows we will use
$$
{\cal L}_{14b} = a_1 \eptwo D_\mu \Omega_\nu{}^{cd} B^{ac} B^{bd}
$$

\noindent
{\bf Vertex $2-2-1$ with three derivatives} Here the most general
ansatz has the form:
\begin{eqnarray*}
{\cal L}_{13} &=& \epthree [ b_1 D_\mu \Omega_\nu{}^{ab} B^{cd}
f_\alpha{}^d + b_2 D_\mu \Omega_\nu{}^{ad} B^{bd} f_\alpha{}^c + b_3
D_\mu \Omega_\nu{}^{ad} B^{bc} f_\alpha{}^d ] \\
 && + \eptwo [ b_4 \Omega_\mu{}^{ab} \Omega_\nu{}^{cd} B^{cd} + b_5
\Omega_\mu{}^{ac} \Omega_\nu{}^{cd} B^{bd} ]
\end{eqnarray*}
First of all note that in this case we have one possible field
redefinition
$$
f_\mu{}^a \Rightarrow f_\mu{}^a + \kappa_4 B^{ab} f_\mu{}^b
$$
and one on-shell identity (again up to the lower derivative terms)
$$
0 \approx \epfour D_\mu \Omega_{\nu,\alpha}{}^a B^{bc} f_\beta{}^d =
\epthree D_\mu \Omega_\nu{}^{ad} [ B^{bc} f_\alpha{}^d - 2 B^{bd}
f_\alpha{}^c ]
$$
Moreover, it is easy to check that invariance under the $\xi^a$
transformations requires $b_2 = - 2b_3$, so the terms with the
coefficients $b_{2,3}$ vanish on-shell, while the one with coefficient
$b_1$ can be removed by field redefinition. 

Collecting all things together let us consider the following cubic
Lagrangian:
\begin{eqnarray}
{\cal L}_1 &=& a_0 \epfour D_\mu \Omega_\nu{}^{ab}
\Omega_\alpha{}^{ce} \Omega_\beta{}^{de} + a_1 \eptwo D_\mu
\Omega_\nu{}^{cd} B^{ac} B^{bd} \nonumber \\
 && + \eptwo [ b_4 \Omega_\mu{}^{ab} \Omega_\nu{}^{cd} B^{cd} + b_5
\Omega_\mu{}^{ac} \Omega_\nu{}^{cd} B^{bd} ]
\end{eqnarray}
$\eta^{ab}$ transformations produce the following variations for this
Lagrangian:
\begin{eqnarray*}
\delta_0 {\cal L}_1 &=& \frac{2m(d-4)a_0}{(d-2)} \epthree [ D_\mu
\Omega_\nu{}^{ab} B^{cd} \eta^{cd} - 4 D_\mu \Omega_\nu{}^{ac} B^{cd}
\eta^{bd} ] \\
 && + \epthree D_\mu \Omega_\nu{}^{cd} [ \frac{2m(d-4)a_0}{(d-2)}
B^{cd} \eta^{ab} - 2ma_1 B^{ac} \eta^{bd} ] \\
 && + \eptwo [ b_4 D_\mu B^{cd} ( \Omega_\nu{}^{ab} \eta^{cd} -
\Omega_\nu{}^{cd} \eta^{ab} ) + b_5 D_\mu B^{ac} ( \Omega_\nu{}^{bd}
\eta^{cd} - \Omega_\nu{}^{cd} \eta^{bd} ] \\
 && - m \eptwo [ b_4 \Omega_\mu{}^{ab} \Omega_\nu{}^{cd} \eta^{cd} +
b_5 \Omega_\mu{}^{ac} \Omega_\nu{}^{cd} \eta^{bd} ]
\end{eqnarray*}
The terms in the first line can be compensated by the following
corrections to gauge transformations:
$$
\delta_1 f_\mu{}^a = \alpha_1 B^{ab} \eta_\mu{}^b + \alpha_2 \eta^{ab}
B_\mu{}^b + \alpha_3 e_\mu{}^a (B\eta)
$$
while for the second line we use on-shell identity
$$
0 \approx D_\mu \Omega_{\nu,\alpha}{}^d B^{ad} \eta^{bc} = \epthree
D_\mu \Omega_\nu{}^{cd} [ B^{cd} \eta^{ab} - 2 B^{ac} \eta^{bd} ]
$$
and obtain
$$
a_1 = \frac{2(d-4)a_0}{(d-2)}
$$
The remaining terms cannot be compensated by any corrections to gauge
transformations so we have to put
$$
b_4 = b_5 = 0
$$
Thus we get rather simple vertex with four derivatives. But such
vertex exists in $d > 4$ dimensions only, while it is well known that
in $d=4$ there exists cubic vertex having no more that two
derivatives. So we proceed and consider the following ansatz:
\begin{eqnarray}
{\cal L}_1 &=& c_1 \epthree \Omega_\mu{}^{ad} \Omega_\nu{}^{bd}
f_\alpha{}^c + c_2 \epfour D_\mu \Omega_\nu{}^{ab} f_\alpha{}^c
f_\beta{}^d \nonumber \\
 && + c_3 e^\mu{}_a B^2 f_\mu{}^a + c_4 \epthree f_\mu{}^a B^{bc}
D_\nu B_\alpha \nonumber \\
 && + d_1 \epthree \Omega_\mu{}^{ab} B_\nu f_\alpha{}^c + d_2 \eptwo
f_\mu{}^a B^{bc} f_\nu{}^c
\end{eqnarray}

\noindent
{\bf $\xi^a$ transformations} produce the following variations:
$$
\epthree [ - (2c_1+4c_2) D_\mu \Omega_\nu{}^{ad}
\Omega_\alpha{}^{bd} \xi^c - 2c_2 D_\mu \Omega_\nu{}^{ab}
\Omega_\alpha{}^{cd} \xi^d ] - (\frac{4m(d-3)c_2}{(d-2)}+d_1) \epthree
D_\mu \Omega_\nu{}^{ab} B_\alpha \xi^c
$$
$$
+ e^\mu{}^a [ - (2c_3+c_4) B^{bc} D_\mu B^{bc} \xi^a - 2c_4 D_\mu
B^{ab} B^{bc} \xi^c ] + (d_2+mc_4) \eptwo D_\mu B^{ac} [ f_\nu{}^c
\xi^b - f_\nu{}^b \xi^c ] 
$$
$$
+ e^\mu{}_a [ (d_1-d_2) \Omega_\mu{}^{bc} B^{bc} \xi^a + (mc_4-d_2)
B^{ab} \Omega_\mu{}^{bc} \xi^c + (2d_1-d_2-mc_4) \Omega_\mu{}^{ab}
B^{bc} \xi^c ] 
$$
$$
+ (\frac{4m^2(d-3)c_2}{(d-2)}+md_1) \eptwo \Omega_\mu{}^{ac} [
f_\nu{}^b \xi^c - f_\nu{}^c \xi^b ] - 2m(d_2+mc_4) e^\mu{}_a B^{ab}
\xi^b B_\mu
$$
If we put
\begin{equation}
c_1 = - 2c_2, \qquad c_3 = - \frac{c_4}{2}, \qquad d_1 = d_2 = - mc_4
\end{equation}
we obtain
$$
- c_2 \epthree {\cal F}_{\mu\nu}{}^{ab} \Omega_\alpha{}^{cd} \xi^d -
2c_4 e^\mu{}_a {\cal B}_\mu{}^{ab} B^{bc} \xi^c
+ 2m(c_4-2c_2) e^\mu{}_a B^{ab} \Omega_\mu{}^{bc} \xi^c
$$
$$
- (\frac{4m(d-3)c_2}{(d-2)}-mc_4) \epthree D_\mu \Omega_\nu{}^{ab}
B_\alpha \xi^c + (\frac{4m^2(d-3)c_2}{(d-2)}-m^2c_4) \eptwo
\Omega_\mu{}^{ac} [ f_\nu{}^b \xi^c - f_\nu{}^c \xi^b ] 
$$
First two terms can be compensated by the following corrections to
gauge transformations:
$$
\delta_1 f_\mu{}^a \sim \Omega_\mu{}^{ab} \xi^b, \qquad
\delta_1 B_\mu \sim B^{ab} \xi^b
$$
while  the remaining terms require
$$
c_4 = 2c_2, \qquad c_4 = \frac{4(d-3)c_2}{(d-2)} \quad \Leftrightarrow
\quad d=4
$$
thus such solution indeed exists in $d=4$ dimensions only.

\noindent
{\bf $\eta^{ab}$ transformations} With the same restrictions on
parameters we get
$$
c_2 \epthree {\cal F}_{\mu\nu}{}^{ab} \eta^{cd} f_\alpha{}^d - 4m
(c_4 -2c_2) \eptwo \Omega_\mu{}^{ac} \eta^{bc} B_\nu
$$
$$
- 2m(c_4-2c_2) e^\mu{}_a B^{ab} \eta^{bc} f_\mu{}^c - m^2
(c_4 - \frac{4(d-3)c_2}{(d-2)}) \eptwo f_\mu{}^a \eta^{bc} f_\nu{}^c
$$
The first term can be compensated by the following correction
$$
\delta f_\mu{}^a \sim \eta^{ab} f_\mu{}^b
$$
while the remaining ones again give
$$
c_4 = 2c_2, \qquad d = 4
$$
\noindent
{\bf $\xi$ transformations} Similarly:
$$
m(-c_4+\frac{4(d-3)c_2}{(d-2)}) \epthree D_\mu \Omega_\nu{}^{ab}
f_\alpha{}^c + 2m(-2c_2+c_4) \eptwo \Omega_\mu{}^{ac}
\Omega_\nu{}^{bc}
 + \frac{m(d-4)c_4}{(d-2)} B^2 
$$
in agreement with all previous results.

\subsection{Gravitational interaction}

In this case we have to consider variations for all five
transformations: $\hat{\eta}^{ab}$, $\hat{\xi}^a$ for graviton and
$\eta^{ab}$, $\xi^a$, $\xi$ for partially massless spin-2. It requires
rather long calculations so we will not reproduce it here restricting
ourselves with the main results.

\noindent
{\bf Vertices with four derivatives} Using on-shell identities and
field redefinitions can be written as follows
$$
{\cal L}_{14} = a_0 \epfour [ 2 D_\mu \Omega_\nu{}^{ab}
\Omega_\alpha{}^{ce} \omega_\beta{}^{de} + D_\mu \omega_\nu{}^{ab}
\Omega_\alpha{}^{ce} \Omega_\beta{}^{de} ] + a_1 \eptwo D_\mu
\omega_\nu{}^{cd} B^{ac} B^{bd}
$$

\noindent
{\bf Vertex with three derivatives} As in the case of
self-interaction all terms of the form $D\Omega Bh$ and $D\omega Bf$
vanish on-shell or can be removed by field redefinitions. This leaves
us with:
$$
{\cal L}_{13} = \eptwo [ b_1 \Omega_\mu{}^{ab} B^{cd}
\omega_\nu{}^{cd} + b_2 \Omega_\mu{}^{ac} B^{cd} \omega_\nu{}^{bd} +
b_3 \Omega_\mu{}^{cd} B^{cd} \omega_\nu{}^{ab} + b_4 \Omega_\mu{}^{cd}
B^{ab} \omega_\nu{}^{cd} ]
$$
Note that this structure is similar to one of $2-2-1$ vertex with
three derivatives that plays important role in the electromagnetic
interactions for spin 2 particles \cite{Zin09}.

\noindent
{\bf Variations of order $m$} require\footnote{We organize variations
by the dimensionality of coefficients. E.g. variations of order $m$
means coefficients of the form $ma$ or $b$ and so on.}: 
$$
b_2 = - 4b_1, \qquad b_3 = b_1, \qquad b_1 + b_4 = - 
\frac{2m(d-4)a_0}{(d-2)}, \qquad a_1 = \frac{4(d-4)a_0}{(d-2)}
$$
provided we introduce the following corrections to the gauge
transformations:
$$
\delta B_\mu = b_4 [ \omega_\mu{}^{ab} \eta^{ab} - \Omega_\mu{}^{ab}
\hat{\eta}^{ab} ]
$$
Thus at this stage we have two independent parameters $a_0$ and say
$b_4$.

\noindent
{\bf Vertices with two derivatives} The most general on-shell
non-equivalent form looks like
\begin{eqnarray}
{\cal L}_{12} &=& \epthree [ c_1 \Omega_\mu{}^{ad} \Omega_\nu{}^{bd}
h_\alpha{}^c + c_2 \Omega_\mu{}^{ab} \Omega_\nu{}^{cd} h_\alpha{}^c ]
\nonumber \\
 && + \epthree [ c_3 \Omega_\mu{}^{ad} \omega_\nu{}^{bd} f_\alpha{}^c
+ c_4 \Omega_\mu{}^{ab} \omega_\nu{}^{cd} f_\alpha{}^d + c_5
\Omega_\mu{}^{ad} \omega_\nu{}^{bc} f_\alpha{}^d ] \nonumber \\
 && + e^\mu{}_a [ c_6 B^2 h_\mu{}^a + c_7 B^{ab} B^{bc} h_\mu{}^c ]
\end{eqnarray}

\noindent
{\bf Variations of order $m^2$} require
$$
c_1 + c_5 = - 2mb_1, \qquad c_2 + c_5 = m(b_4-b_1)
$$
$$
c_3 = - 2c_5, \qquad c_4 = - c_4, \qquad 
c_7 = 4c_6 - \frac{2ma_1}{(d-2)}
$$
while corresponding corrections to the gauge transformations gave the
form:
\begin{eqnarray}
\delta f_\mu{}^a &=& - 2c_5 (\hat{\eta}^{ab} f_\mu{}^b -
\omega_\mu{}^{ab} \xi^b) +  2(c_1 - \frac{4m^2(d-4)a_0}{(d-2)})
(\eta^{ab} h_\mu{}^b - \Omega_\mu{}^{ab} \hat{\xi}^b) \nonumber \\
\delta h_\mu{}^a &=& 2c_5 (\eta^{ab} f_\mu{}^b - \Omega_\mu{}^{ab}
\xi^b), \qquad \delta B_\mu = 2c_6 B_\mu{}^a \hat{\xi}^a 
\end{eqnarray}

\noindent
{\bf Vertex with one derivative} The most general ansatz is:
$$
{\cal L}_{11} = \epthree [ d_1 \omega_\mu{}^{ab} f_\nu{}^c B_\alpha +
d_2 \Omega_\mu{}^{ab} h_\nu{}^c B_\alpha ] + d_3 \eptwo B^{ab}
f_\mu{}^c h_\nu{}^c
$$
Note that the only possible term without derivatives
$$
\epthree f_\mu{}^a f_\nu{}^b h_\alpha{}^c
$$
is forbidden by the invariance under the $\xi$ transformations.

First of all note that solution with non-zero parameter $c_5$ exists
in $d=3$ dimensions only. Recall that the Fradkin-Vasiliev formalism
we use in the main part works in $d \ge 4$ dimensions only so it 
cannot reproduce such vertex. Note however that in the frame-like
gauge invariant formalism this vertex (having no more than two
derivatives) has been constructed in \cite{Zin12}. For the general $d
> 3$ case we obtain two independent solutions with $a_0$ and $b_4$ as
free parameters:
$$
c_1 = - 2mb_1, \qquad c_2 = m(b_4-b_1), \qquad 
c_6 = \frac{m(d-3)b_4}{(d-2)}
$$
$$
c_3 = c_4 = c_5 = d_1 = d_2 = d_3 = 0
$$
Note at last that in $d=4$ dimensions parameter $a_0$ is absent
leaving us with one vertex only, moreover in this case all four
derivative terms vanish on-shell.

\end{document}